\documentclass[aps,pra,preprintnumbers,showpacs,tightenlines]{revtex4}
\usepackage{amssymb}
\usepackage{amsmath}
\usepackage{graphicx}
\usepackage{epsfig}
\usepackage{subfigure}
\usepackage{amsfonts}
\usepackage{CJK}

\begin{document}

\title{Entangling superconducting qubits in a multi-cavity system}


\author{Chui-Ping Yang$^{1,2,3}$, Qi-Ping Su$^{1,2,3}$, Shi-Biao Zheng $^{4}$, and Franco Nori$^{1,2}$}
\address{$^1$CEMS, RIKEN, Saitama 351-0198, Japan}
\address{$^2$Department of Physics, The University of Michigan, Ann Arbor, Michigan 48109-1040, USA}
\address{$^3$Department of Physics, Hangzhou Normal University, Hangzhou, Zhejiang 310036, China}
\address{$^4$Department of Physics, Fuzhou University, Fuzhou 350108, China}
\date{\today}

\begin{abstract}
Important tasks in cavity quantum electrodynamics include the generation and control of
quantum states of spatially-separated particles distributed in different cavities.
An interesting question in this context is how to prepare entanglement among particles located in different cavities,
which are important for large-scale quantum information processing. We here
consider a multi-cavity system where cavities are coupled to a superconducting (SC) qubit
and each cavity hosts many SC qubits. We show that all intra-cavity SC qubits plus the coupler SC qubit can be prepared in
an entangled Greenberger-Horne-Zeilinger (GHZ) state, by using a
single operation and without the need of measurements. The GHZ state is
created without exciting the cavity modes; thus greatly suppressing the
decoherence caused by the cavity-photon decay and the effect of unwanted
inter-cavity crosstalk on the operation. We also introduce two simple
methods for entangling the intra-cavity SC qubits in a GHZ state. As an example,
our numerical simulations show that it is feasible, with current circuit-QED
technology, to prepare high-fidelity GHZ states, for up to nine SC qubits by using SC qubits
distributed in two cavities. This proposal can in principle be used to implement
a GHZ state for {\it an arbitrary number} of SC qubits distributed in multiple cavities. The proposal is quite general
and can be applied to a wide range of physical systems, with the intra-cavity qubits being either atoms, NV
centers, quantum dots, or various SC qubits.
\end{abstract}

 \maketitle
\date{\today}

\section{Introduction}

Superconducting devices can be fabricated using modern integrated circuit
technology, their properties can be characterized and adjusted in situ, and
their coherence time has recently been significantly increased [1-9].
Moreover, various single- and multiple-qubit operations with state readout
have been demonstrated [10-15], and nonlinear optical processes in a
superconducting quantum circuit have been investigated [16]. In addition,
Circuit QED, consisting of microwave resonators and SC qubits, is
particularly attractive and considered as one of the leading candidates for
QIP [17-23]. The strong and ultrastrong coupling between a microwave cavity
and SC qubits has been demonstrated in experiments (e.g., [24-26]). In
addition, using SC qubits coupled to a \textit{single} cavity or resonator
(hereafter, the terms cavity and resonator are used interchangeably), a
number of theoretical proposals have been presented for realizing quantum
gates and entanglement [17-19,27-32], and two- and three-qubit quantum gates
and three-qubit entanglement have been experimentally demonstrated [33-37].

In recent years, there is much interest in large-scale QIP, which usually
involves many qubits. Note that placing all qubits in a single cavity may
cause many problems, such as increasing the cavity decay rate and decreasing
the qubit-cavity coupling strength. Therefore, for cavity or circuit
QED-based large-scale QIP, the qubits should be distributed in different
cavities, and the ability to perform nonlocal quantum operations for these
qubits is a prerequisite to realize distributed quantum computation. During
the past few years, attention has been paid to the preparation of entangled
states of two or more cavities, or of qubits located in different cavities,
and implementation of quantum logic gates on photons/qubits distributed over
different cavities in a network. Specifically, rapid progress has been
achieved in the following two directions:

(i) Manipulating and generating nonclassical microwave field states with
photons distributed in different cavities. By using a SC qubit (artificial
atom) coupled to cavities, schemes have been proposed for synthesizing
different entangled photonic states of two SC resonators [38], and for
generating multi-particle entangled states of photons in different cavities
[39,40]. By employing the idea of Ref.~[41] the so-called NOON state of
photons in two resonators has been experimentally created [42]. In addition,
how to perform quantum logic operations on photons located in different
cavities has been investigated [43].

(ii) Quantum state engineering and quantum operations with qubits
distributed in different cavities. By using a SC qubit to couple two or more
cavities/resonators, proposals have been presented for generating GHZ states
with multiple SC qubits coupled to multiple resonators via employing cavity
photons and through step-by-step control [40,44], and for quantum
information transfer between two spatially-separated SC qubits distributed
in two cavities [45]. Recently, GHZ states of three SC qubits in circuits
consisting of two resonators have been experimentally prepared [7].
Furthermore, using an intermediate SC qubit coupled to two planar
resonators, quantum teleportation between two distant SC qubits has recently
been demonstrated in experiments [46].

GHZ states are not only of great interest for fundamental tests of quantum
mechanics [47], but also have applications in QIP [48], quantum
communications [49], error-correction protocols [50], quantum metrology
[51], and high-precision spectroscopy [52]. During the past years,
experimental realizations of GHZ states with eight photons using linear
optical devices [53,54], fourteen ions [55], three SC qubits in circuit QED
[7], five SC qubits via capacitance coupling [56], and three qubits in NMR
[57] have been reported. Theoretically, proposals for generating entangled
states with SC qubit circuits have been presented [58]. In addition, based
on cavity QED or circuit QED, a large number of theoretical methods have
been presented for creating multi-qubit GHZ states with various physical
systems (e.g., atoms, quantum dots, and SC devices) that are coupled to a
single cavity/resonator mode [59-68]. However, we note that how to generate
GHZ states with qubits in different cavities has not been thoroughly
investigated.

Motivated by the above, here we present an efficient method to entangle SC
qubits in a multi-cavity system, where cavities are coupled to a SC qubit
and each cavity hosts many SC qubits. We show that the cavity-induced
effective conditional dynamics between the intra-cavity SC qubits and the
coupler SC qubit can be employed to entangle all the SC qubits in a GHZ
state. In this work, we also introduce two simple methods for entangling the
intra-cavity SC qubits in a GHZ state. As an example, our numerical
simulations show that it is feasible, with current circuit-QED technology,
to prepare high-fidelity GHZ states, for up to nine SC qubits by using SC
qubits embedded in two cavities. To the best of our knowledge, based on
circuit QED, the experimental demonstration of GHZ states has only been
reported for three SC qubits [7,34].

This proposal has the following advantages: (i) The GHZ state preparation
does not require step-by-step control, which involves only one operation for
entangling all qubits and a few basic operations for entangling the
intra-cavity qubits; (ii) The entanglement is prepared without exciting the
cavity photons, and thus the decoherence induced by cavity decay and the
effect of unwanted inter-cavity crosstalk on the operation are greatly
suppressed; (iii) Because none of the intra-cavity qubits is excited during
the operation, decoherence from the qubits is much reduced; (iv) More
interestingly, this proposal can in principle be used to implement a GHZ
state for \textit{an arbitrary number }of qubits distributed in multiple
cavities by using a single coupler qubit, which is important for the future
realization of large-scale QIP; and (v) We further stress that this proposal
is quite general and can be used for other kinds of qubits, such as atoms,
NV centers, and quantum dots.

This paper is organized as follows. In Sec.~2, we introduce the physical
model considered in this work and derive the effective Hamiltonian used for
the entanglement production. In Sec.~3, we show how to generate GHZ states
for all the intra-cavity SC qubits and the coupler SC qubit based on the
effective Hamiltonian. In Sec.~4, we further introduce two simple methods
for generating GHZ states of intra-cavity SC qubits. In Sec.~5, as an
example, we numerically analyze the experimental feasibility of preparing a
GHZ state of up to nine SC qubits with SC qubits distributed in two
cavities. A concluding summary is given in Sec.~6. For the numerical
calculations, here we use the QuTiP software [69,70].

\section{Physical model and effective Hamiltonian}

\begin{figure}[tbp]
\begin{center}
\includegraphics[bb=162 322 465 595, width=6.5 cm, clip]{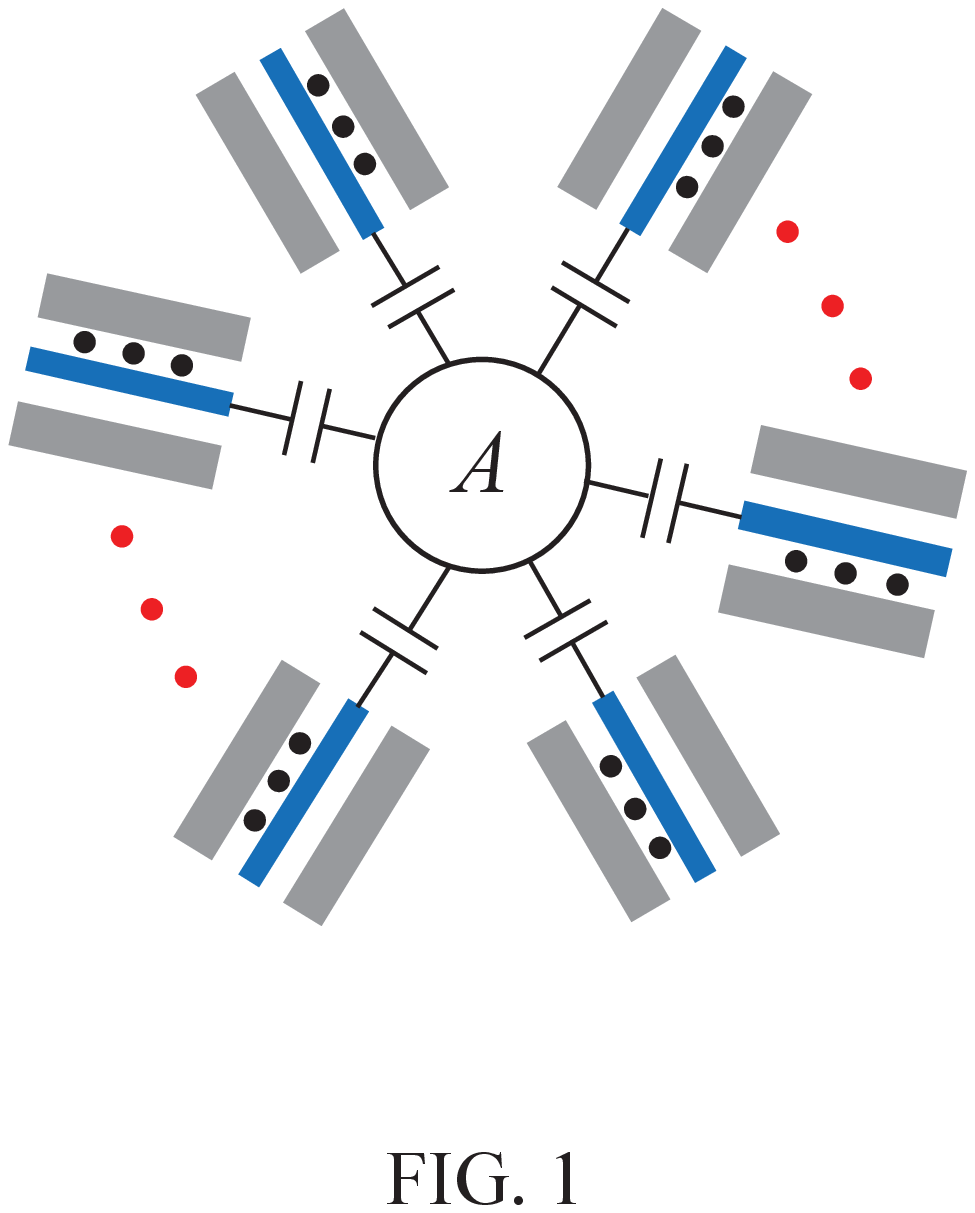} \vspace*{%
-0.08in}
\end{center}
\caption{(Color online) Diagram of a coupler qubit $A$ (circle at the
center) and $N$ cavities ($1,2,...,N$) each hosting qubits. Each cavity here
is a one-dimensional coplanar waveguide transmission line resonator. The
circle $A$ represents a SC qubit, which is capacitively coupled to each
cavity. A dark dot represents an intra-cavity SC qubit. For simplicity, only
three qubits are drawn in each cavity.}
\label{fig:1}
\end{figure}

\begin{figure}[tbp]
\begin{center}
\includegraphics[bb=64 184 457 414, width=10.0 cm, clip]{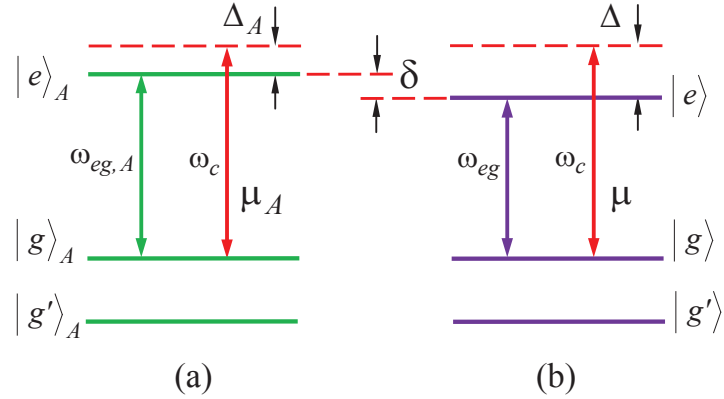} \vspace*{%
-0.08in}
\end{center}
\caption{(Color online) (a) Each cavity is dispersively coupled to the $%
\left\vert g\right\rangle \leftrightarrow \left\vert e\right\rangle $
transition of the coupler qubit $A$, with coupling strength $\protect\mu_A$
and detuning $\Delta_A=\protect\omega_c-\protect\omega_{eg,A}$. Here, $%
\protect\omega_c$ is the cavity frequency while $\protect\omega_{eg,A}$ is
the $\left\vert g\right\rangle \leftrightarrow \left\vert e\right\rangle $
transition frequency of the coupler qubit $A$. (b) Each cavity is
dispersively coupled to its hosting intra-cavity qubits' $\left\vert
g\right\rangle \leftrightarrow \left\vert e\right\rangle $ transition, with
coupling strength $\protect\mu$ and detuning $\Delta=\protect\omega_c-%
\protect\omega_{eg}$. Here, $\protect\omega_{eg}$ is the $\left\vert
g\right\rangle \leftrightarrow \left\vert e\right\rangle $ transition
frequency of the intra-cavity qubits. In addition, $\protect\delta%
=\Delta-\Delta_A$. In (a) and (b), a horizontal solid line represents an
energy level of a qubit; and each vertical (red) double-arrow line, linked
to a middle solid line and a top (red) dashed line, represents the cavity
frequency $\protect\omega_c$. For simplicity, we here consider the case when
the level spacing between the two lowest levels is smaller than that between
the upper two levels. This type of level structure is available in
superconducting charge qubits or flux qubits. Alternatively, the level
spacing between the two lowest levels can be larger than that between the
upper two levels, which applies to superconducting phase, transmon or Xmon
qubits. In (a) and (b), the ground level is labeled by $\left\vert
g^{\prime}_A\right\rangle$ ($\left\vert g^{\prime}\right\rangle$) and the
first excited level is denoted as $\left\vert g_A\right\rangle$ ($\left\vert
g\right\rangle$). One could also denote the ground level as $\left\vert
g_A\right\rangle$ ($\left\vert g\right\rangle$) and the first excited level
as $\left\vert g^{\prime}_A\right\rangle$ ($\left\vert
g^{\prime}\right\rangle$).}
\label{fig:2}
\end{figure}

We consider a system composed of $N$ cavities and assume that cavity $j$\
hosts $m_j$\ SC qubits denoted as $j_1,$\ $j_2,...,$\ and $j_{m_j}$. These
cavities ($1,2,...,N$) are coupled to a common SC qubit $A$\ (coupler
qubit), as shown in Fig.~1. Each qubit\ considered here has three levels,
which are denoted as $\left| g\right\rangle ,$\ $\left| g^{\prime
}\right\rangle ,$\ and $\left| e\right\rangle $\ (Fig.~2). The two logical
states of each intra-cavity qubit are represented by the two levels $\left|
g\right\rangle $ and\ $\left| g^{\prime }\right\rangle $, while those of the
coupler qubit are represented by the two levels $\left| g^{\prime
}\right\rangle $ and\ $\left| e\right\rangle .$ The third level $\left|
e\right\rangle $ for each intra-cavity qubit or $\left| g\right\rangle $ for
the coupler qubit acts as an auxiliary level for realizing, e.g., a
conditional phase shift. The level spacing of the coupler qubit $A$\ is
different from those of the intra-cavity qubits.\textbf{\ }The $\left|
g\right\rangle \leftrightarrow \left| e\right\rangle $ transition of the
intra-cavity qubits is coupled to their respective cavities with coupling
strength $\mu ,$ while the $\left| g\right\rangle \leftrightarrow \left|
e\right\rangle $ transition of the coupler qubit $A$ is coupled to all the
cavities with coupling strength $\mu _A$ (Fig.~2). We here assume that the
level $\left| g^{\prime }\right\rangle $ of each qubit is not affected
during the operation, which can be a good approximation when the transitions
between $\left| g^{\prime }\right\rangle $ and other levels are sufficiently
weak or the relevant transition frequencies are highly detuned from the
cavity frequency. The level spacings of superconducting qubits can be
rapidly adjusted by varying the external control parameters (e.g., the
magnetic flux applied to a superconducting loop for phase, transmon or flux
qubits; see e.g. [20,71-73]), so that the transitions associated with $%
\left| g^{\prime }\right\rangle $ can be tuned far off-resonance with the
resonators. In the interaction picture with respect to the free Hamiltonian
of the system (not shown for simplicity), the Hamiltonian is given by
\begin{equation}
H=\sum_{j=1}^N\mu \left( e^{-i\Delta t}a_jJ_j^{+}+e^{i\Delta t}a_j^{\dagger
}J_j\right) +\sum_{j=1}^N\mu _A\left( e^{-i\Delta _At}a_j\sigma
_A^{+}+e^{i\Delta _At}a_j^{\dagger }\sigma _A\right) ,
\end{equation}
where $J_j^{+}=\sum_{i=1}^{m_j}\sigma _{j_i}^{+},$ $J_j=\sum_{i=1}^{m_j}%
\sigma _{j_i},$ $\sigma _{j_i}^{+}=\left| e\right\rangle _{j_i}\left\langle
g\right| ,$\ $\sigma _{j_i}=\left| g\right\rangle _{j_i}\left\langle
e\right| ,\sigma _A^{+}=\left| e\right\rangle _A\left\langle g\right| ,$\ $%
\sigma _A=\left| g\right\rangle _A\left\langle e\right| ,$ $\Delta =\omega
_c-\omega _{eg},$ and $\Delta _A=\omega _c-\omega _{eg,A}.$ Here, $\omega _c$%
\ is the cavity frequency; and $\omega _{eg}$\ and $\omega _{eg,A}$\ are the
$\left| g\right\rangle \leftrightarrow \left| e\right\rangle $\ transition
frequencies for the intra-cavity qubits and the coupler qubit $A$,
respectively (Fig.~2).

Under the large-detuning condition $\Delta ,\Delta _A,\gg \mu ,\mu _A$, $%
\left| \delta \right| =\left| \Delta -\Delta _A\right| ,$ the dynamics
governed by $H$\ is equivalent to that decided by the following effective
Hamiltonian [74-76]
\begin{eqnarray}
H_{\mathrm{eff}} &=&\frac{\mu ^2}\Delta \sum_{j=1}^N\left( {G}%
_ja_j^{+}a_j-E_ja_ja_j^{+}\right)  \nonumber \\
&&\ \ +\frac{\mu _A^2}{\Delta _A}\sum_{j=1}^N\left( {G}%
_Aa_j^{+}a_j-E_Aa_ja_j^{+}\right)  \nonumber \\
&&\ \ -\frac{\mu ^2}\Delta \sum_{j=1}^N\left( J_jJ_j^{+}-{G}_j\right)
\nonumber \\
&&\ \ -\lambda \sum_{j=1}^N(e^{-i\delta t}J_j\sigma _A^{+}+H.c.)  \nonumber
\\
&&\ \ +\sum_{j\neq k=1}^N\lambda _{jk}\left( a_ja_k^{+}+H.c.\right) \left(
E_A-G_A\right) ,
\end{eqnarray}
where\textbf{\ }$\delta =\Delta -\Delta _A,$ ${G}_j=\sum_{i=1}^{m_j}\left|
g\right\rangle _{j_i}\left\langle g\right| ,$\textbf{\ }${E}%
_j=\sum_{i=1}^{m_j}\left| e\right\rangle _{j_i}\left\langle e\right| ,$%
\textbf{\ }${G}_A=\left| g\right\rangle _A\left\langle g\right| ,$\textbf{\ }%
${E}_A=\left| e\right\rangle _A\left\langle e\right| ,$\textbf{\ }$\lambda
_{jk}=\mu _A^2/\Delta _A,$ and\textbf{\ }$\lambda =\frac{\mu \mu _A}2\left(
\frac 1\Delta +\frac 1{\Delta _A}\right) .$\textbf{\ }The terms in the first
and second lines of (2) account for the ac-Stark shifts of the level $\left|
g\right\rangle $\ ($\left| e\right\rangle $) of the intra-cavity qubits and
the coupler qubit $A$\ induced by the corresponding cavity modes,
respectively. In addition, the terms in the third line describe the\emph{\ }%
effective dipole-dipole interaction between the intra-cavity qubits located
in the same cavities. The terms in the fourth line describe the
dipole-dipole interaction between the intra-cavity qubits and the coupler
qubit $A$, and the terms in the last (fifth) line characterize the coupling
between any two cavities.

If each cavity is initially in a vacuum state, the Hamiltonian (2) reduces to%
\textbf{\ }
\begin{eqnarray}
H_{\mathrm{eff}} &=&-\frac{\mu ^{2}}{\Delta }\sum_{j=1}^{N}E_{j}-\frac{N\mu
_{A}^{2}}{\Delta _{A}}E_{A}  \nonumber \\
&&\ \ -\frac{\mu ^{2}}{\Delta }\sum_{j=1}^{N}\left( J_{j}J_{j}^{+}-{G}%
_{j}\right)   \nonumber \\
&&\ \ -\lambda \sum_{j=1}^{N}(e^{-i\delta t}J_{j}\sigma _{A}^{+}+H.c.).
\end{eqnarray}%
According to [74], if\textbf{\ }$\left\vert \delta \right\vert \gg \lambda ,$%
\textbf{\ }$\frac{N\mu ^{2}}{\Delta }$\textbf{, }$\frac{N\mu _{A}^{2}}{%
\Delta _{A}}$\textbf{,} the terms in the last line can be\ replaced by\ $%
\frac{\lambda ^{2}}{\delta }[\sum_{j=1}^{N}J_{j}\sigma
_{A}^{+},\sum_{j=1}^{N}J_{j}^{+}\sigma _{A}]$. Thus, the dynamics governed
by Hamiltonian~(3) is approximately equivalent to that by the following
Hamiltonian\textbf{\ }
\begin{eqnarray}
H_{\mathrm{eff}}^{^{\prime }} &=&-\frac{\mu ^{2}}{\Delta }%
\sum_{j=1}^{N}E_{j}-\frac{N\mu _{A}^{2}}{\Delta _{A}}E_{A}  \nonumber \\
&&\ \ -\frac{\mu ^{2}}{\Delta }\sum_{j=1}^{N}\left( J_{j}J_{j}^{+}-{G}%
_{j}\right)   \nonumber \\
&&\ \ +\frac{\lambda ^{2}}{\delta }E_{A}\sum_{j=1}^{N}J_{j}%
\sum_{l=1}^{N}J_{l}^{+}-\frac{\lambda ^{2}}{\delta }G_{A}%
\sum_{j=1}^{N}J_{j}^{+}\sum_{l=1}^{N}J_{l},
\end{eqnarray}%
which can be rewritten as\textbf{\ }
\begin{eqnarray}
H_{\mathrm{eff}}^{^{\prime }} &=&-\frac{\mu ^{2}}{\Delta }%
\sum_{j=1}^{N}E_{j}-\frac{N\mu _{A}^{2}}{\Delta _{A}}E_{A}  \nonumber \\
&&\ \ -\frac{\mu ^{2}}{\Delta }\sum_{j=1}^{N}\sum_{i\neq k=1}^{m_{j}}\sigma
_{j_{i}}\sigma _{j_{k}}^{+}  \nonumber \\
&&\ \ +\frac{\lambda ^{2}}{\delta }E_{A}\sum_{j=1}^{N}\sum_{l=1}^{N}%
\sum_{i=1}^{m_{j}}\sum_{k=1}^{m_{l}}\sigma _{j_{i}}\sigma _{l_{k}}^{+}
\nonumber \\
&&\ \ -\frac{\lambda ^{2}}{\delta }G_{A}\sum_{j=1}^{N}\sum_{l=1}^{N}%
\sum_{i=1}^{m_{j}}\sum_{k=1}^{m_{l}}\sigma _{j_{i}}^{+}\sigma _{l_{k}}.
\end{eqnarray}%
When the level $\left\vert g\right\rangle $ of the coupler qubit and the
level $\left\vert e\right\rangle $ of\textbf{\ }the intra-cavity qubits are
not populated, the Hamiltonian (5)\ reduces to
\begin{eqnarray}
H_{\mathrm{eff}}^{^{\prime }} &=&-\frac{N\mu _{A}^{2}}{\Delta _{A}}E_{A}+\frac{%
\lambda ^{2}}{\delta }E_{A}\sum_{j=1}^{N}{G}_{j}  \nonumber \\
\  &=&-\frac{N\mu _{A}^{2}}{\Delta _{A}}\left\vert e\right\rangle
_{A}\left\langle e\right\vert +\frac{\lambda ^{2}}{\delta }\left\vert
e\right\rangle _{A}\left\langle e\right\vert \otimes
\sum_{j=1}^{N}\sum_{i=1}^{m_{j}}\left\vert g\right\rangle
_{j_{i}}\left\langle g\right\vert .
\end{eqnarray}%
This effective Hamiltonian can be turned off by tuning the qubit levels in
such a way that the transitions between these levels are highly off-resonant
with the cavities, and hence the qubits are effectively decoupled from the
corresponding cavities.

\section{Entangling intra-cavity qubits and the coupler qubit}

Let us go back to the setup in Fig. 1. Initially, the qubit system is
decoupled from the cavity system, and each cavity is in the vacuum state.%
\textrm{\ }Assume that each intra-cavity SC qubit is in the state $%
\left\vert +\right\rangle =1/\sqrt{2}\left( \left\vert g^{\prime
}\right\rangle +\left\vert g\right\rangle \right) $ and the coupler SC qubit
is in the state $\alpha \left\vert g^{\prime }\right\rangle _{A}+\beta
\left\vert e\right\rangle _{A}$ ($\left\vert \alpha \right\vert
^{2}+\left\vert \beta \right\vert ^{2}=1$). These states can be prepared
from the qubit ground state with classical pulses. For simplicity, let us
consider the case of $\left\vert g\right\rangle $\ being the ground state (a
case applied to the flux qubits considered in Sec. 5, with three levels
illustrated in Fig. 6). The state $\left\vert g\right\rangle $\ can be
transformed to $\left\vert +\right\rangle $\ by applying a $\pi /2$\ pulse
tuned to the $\left\vert g\right\rangle \leftrightarrow \left\vert
g^{^{\prime }}\right\rangle $\ transition\ [29]. The preparation of the
state $\alpha \left\vert g^{\prime }\right\rangle _{A}+\beta \left\vert
e\right\rangle _{A}$\ consists of two steps: (i) apply a $\pi $\ pulse,
tuned to the $\left\vert g\right\rangle _{A}\leftrightarrow \left\vert
e\right\rangle _{A}$\ transition, to flip the state $\left\vert
g\right\rangle _{A}$\ to $\left\vert e\right\rangle _{A}$; (ii)\ employ a
classical pulse to drive the $\left\vert g^{\prime }\right\rangle
_{A}\leftrightarrow \left\vert e\right\rangle _{A}$\ transition, with the
Rabi frequency $\Omega $\ and duration $t$\ satisfying $\alpha =\cos \left(
\Omega t\right) $\ and $\beta =\sin \left( \Omega t\right) .$\

The initial state of the whole qubit system is thus given by
\begin{equation}
\left( \alpha \left| g^{\prime }\right\rangle _A+\beta \left| e\right\rangle
_A\right) \otimes \prod_{j=1}^N\prod_{i=1}^{m_j}\left| +\right\rangle _{j_i}.
\end{equation}

Now adjust the level spacings of qubits, so that the qubit-resonator
coupling is turned on and the dynamics of the qubit system is governed by
the effective Hamiltonian~(6).\emph{\ }One can see that under the
Hamiltonian (6), the state~(7) evolves into
\begin{eqnarray}
&&\ \alpha \left\vert g^{\prime }\right\rangle
_{A}\prod_{j=1}^{N}\prod_{i=1}^{m_{j}}\left\vert +\right\rangle
_{j_{i}}+\beta e^{-iH_{\mathrm{eff}}^{\prime }t}\left\vert e\right\rangle
_{A}\prod_{j=1}^{N}\prod_{i=1}^{m_{j}}\left\vert +\right\rangle _{j_{i}}
\nonumber \\
\  &=&\alpha \left\vert g^{\prime }\right\rangle
_{A}\prod_{j=1}^{N}\prod_{i=1}^{m_{j}}\left\vert +\right\rangle _{j_{i}}
\nonumber \\
&&+\beta \exp \left( iN\mu _{A}^{2}t/\Delta _{A}\right) \left\vert
e\right\rangle _{A}\prod_{j=1}^{N}\prod_{i=1}^{m_{j}}\left[ \left\vert
g^{\prime }\right\rangle _{j_{i}}+\exp \left( -i\lambda ^{2}t/\delta \right)
\left\vert g\right\rangle _{j_{i}}\right] ,
\end{eqnarray}%
where we have used $H_{\mathrm{eff}}^{\prime }\left\vert g^{\prime
}\right\rangle =0.$ Setting
\begin{eqnarray}
\frac{N\mu _{A}^{2}}{\Delta _{A}}t &=&2m\pi , \\
\frac{\lambda ^{2}}{\left\vert \delta \right\vert }t &=&\pi ,
\end{eqnarray}%
where $m$ is an integer, Eq.~(8) can be expressed as
\begin{equation}
\left\vert \mathrm{GHZ}\right\rangle =\alpha \left\vert g^{\prime
}\right\rangle _{A}\prod_{j=1}^{N}\prod_{i=1}^{m_{j}}\left\vert
+\right\rangle _{j_{i}}+\beta \left\vert e\right\rangle
_{A}\prod_{j=1}^{N}\prod_{i=1}^{m_{j}}\left\vert -\right\rangle _{j_{i}},
\end{equation}%
where $\left\vert -\right\rangle =1/\sqrt{2}\left( \left\vert g^{\prime
}\right\rangle -\left\vert g\right\rangle \right) $. Since $\left\vert
-\right\rangle $ is orthogonal to $\left\vert +\right\rangle $, the state
(11) is a multi-particle GHZ entangled state for the coupler qubit and the
qubits distributed in multiple cavities. One can see that the entangled
state preparation here is based on a $\pi $-phase shift on the state $%
\left\vert g\right\rangle $\ of each intracavity qubit conditional upon the
coupler qubit being in the state $\left\vert e\right\rangle _{A}.$ Note that
by applying a classical pulse to the coupler qubit, the states $\left\vert
g^{\prime }\right\rangle _{A}$ and $\left\vert e\right\rangle _{A}$ can be
easily converted into the states $\left\vert +\right\rangle _{A}$ and $%
\left\vert -\right\rangle _{A}$, respectively. The operation sequence for
preparing the GHZ state (11) is illustrated in Fig. 3.

\begin{figure}[tbp]
\begin{center}
\includegraphics[bb=21 50 584 239, width=10.0 cm, clip]{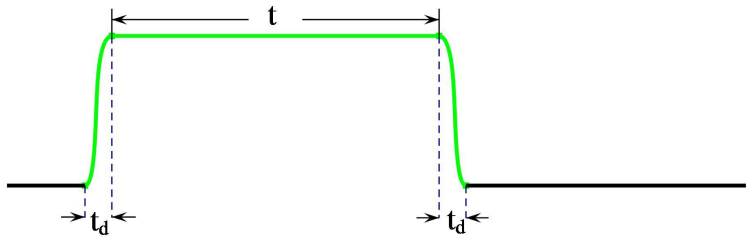} \vspace*{%
-0.08in}
\end{center}
\caption{(Color online) Sequence of operations on each qubit (from left to
right). Here, $t$ satisfies Eqs.(9) and (10), which is the qubit-cavity
interaction time required for producing the GHZ state; while $t_d$ (within
1--3 ns [77,78]) is the time required to adjust the qubit level spacings.
Note that the level spacings of qubits are tuned simultaneously. For
simplicity, here and in Figs. (4) and (5), we assume that the time needed
for adjusting the level spacings is the same for both intra-cavity qubits
and coupler qubit.}
\label{fig:3}
\end{figure}

Note that when the coupling of qubits to the mode in a cavity is spatially
dependent, then different intra-cavity qubits will acquire different
conditional phases. The coupler qubit would acquire a single-qubit phase,
when its coupling to the cavity mode is deviated from the preset value. To
eliminate the effect of this phase, the interaction time should be adjusted
so that this phase is equal to $2m\pi $, with $m$\ being an integer.

It should be mentioned that,\emph{\ }in order to maintain\ the initial
states and the prepared GHZ states of the qubit system, the coupler qubit
and the intra-cavity qubits should be decoupled from their respective
cavities before/after the entanglement production, which requires the
qubit-cavity coupling to be switchable. This requirement can be readily
achieved, by prior adjustment of the level spacings of the qubits [21,71-73]
or the frequencies of the cavities. We note that the rapid tuning of
microwave cavity frequencies has been experimentally demonstrated (e.g., in
less than a few nanoseconds for a superconducting transmission line
resonator [77,78]).

By preparing the initial state of the coupler qubit $A$ with different
values of $\alpha $ and $\beta ,$ the degree of entanglement for the GHZ
state (11) can be adjusted and thus this protocol can be used to generate
GHZ entangled states with \textit{an arbitrary degree of entanglement}. As
shown above, during the entanglement preparation no photons are excited in
each cavity, no measurement is needed, and only a single-step operation is
required.

\section{Entangling intra-cavity qubits}

In this section, we will briefly introduce two methods for preparing all
intra-cavity SC qubits in a GHZ state. The first method requires a
measurement on the coupler SC qubit, while the second one does not need any
measurement.

\begin{center}
\textbf{A. Method 1}
\end{center}

All qubits including the coupler qubit are first prepared in the GHZ
state~(11). One can see from Eq.~(11) that through a unitary tranformation $%
\left| g^{\prime }\right\rangle _A\rightarrow \left( \left| g^{\prime
}\right\rangle _A+\left| e\right\rangle _A\right) /\sqrt{2}$ and $\left|
e\right\rangle _A\rightarrow \left( \left| g^{\prime }\right\rangle
_A-\left| e\right\rangle _A\right) /\sqrt{2}$ and then a measurement on the
coupler qubit $A$, the intra-cavity qubits will be prepared in one of the
two GHZ states $\left| \mathrm{GHZ}\right\rangle ^{\pm }=\alpha
\prod_{j=1}^N\prod_{i=1}^{m_j}\left| +\right\rangle _{j_i}\pm \beta
\prod_{j=1}^N\prod_{i=1}^{m_j}\left| -\right\rangle _{j_i}$ (depending on
the measurement outcome). The operation sequence for preparing the states $%
\left| \mathrm{GHZ}\right\rangle ^{\pm }$ is illustrated in Fig. 4.

\begin{figure}[tbp]
\begin{center}
\includegraphics[bb=7 17 494 320, width=10.0 cm, clip]{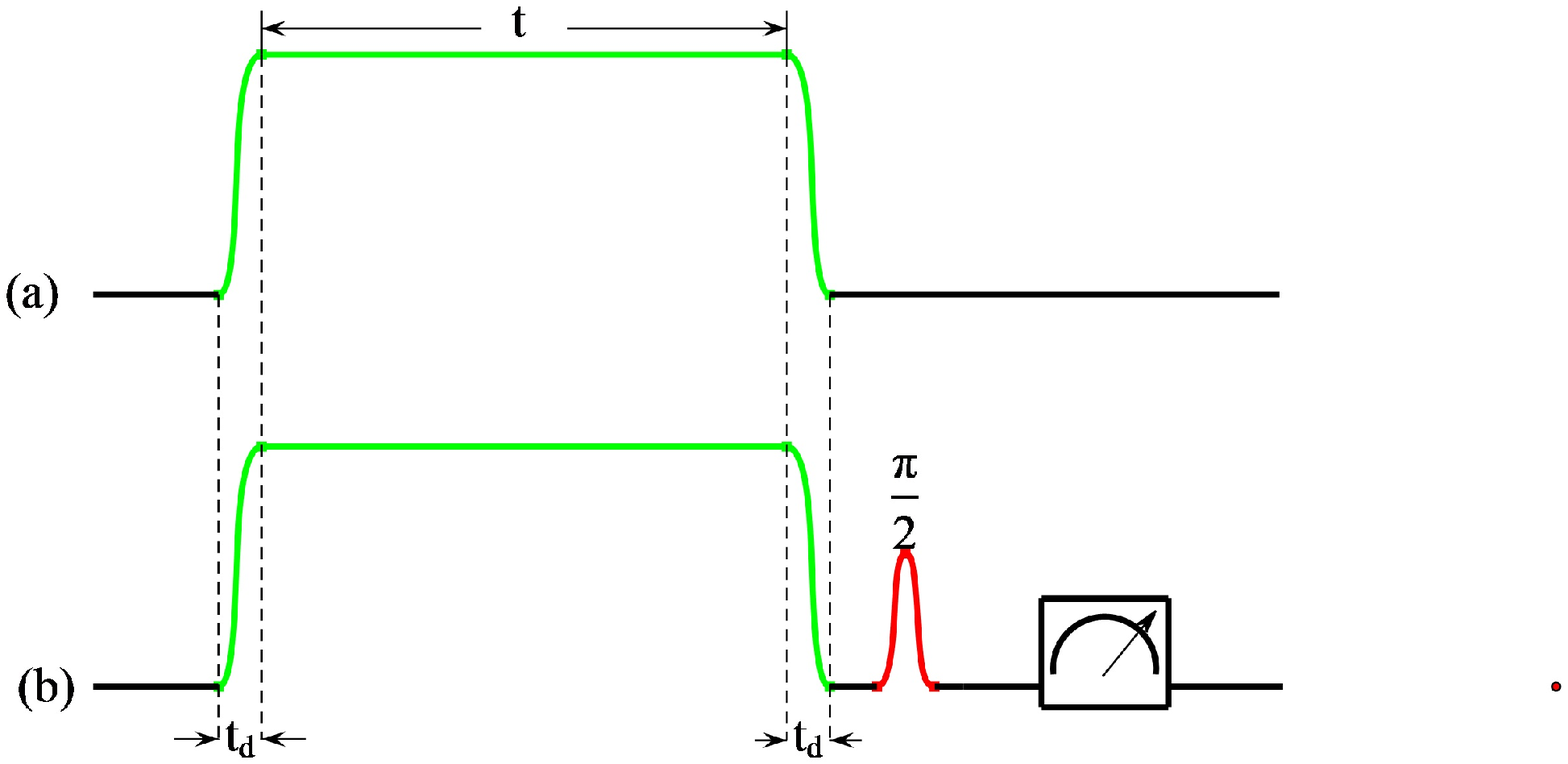} \vspace*{%
-0.08in}
\end{center}
\caption{(Color online) (a) Sequence of operations on each intra-cavity
qubit. (b) Sequence of operations on the coupler qubit. The green curves in
(a) and (b) correspond to the operation for producing the GHZ state (11). In
(b), the $\protect\pi/2$ pulse transforms the state $\left| g^{\prime
}\right\rangle _A\rightarrow \left( \left| g^{\prime }\right\rangle
_A+\left| e\right\rangle _A\right) /\protect\sqrt{2}$ and the state $\left|
e\right\rangle _A\rightarrow \left( \left| g^{\prime }\right\rangle
_A-\left| e\right\rangle _A\right) /\protect\sqrt{2}$, and the square box
with a meter represents a measurement on the coupler qubit along the basis $%
\{\left| g^{\prime }\right\rangle _A$, $\left| e\right\rangle _A\}$. Note
that the operations in (a) and (b) are performed from left to right.}
\label{fig:4}
\end{figure}

The two GHZ states\textbf{\ }$\left| \mathrm{GHZ}\right\rangle ^{+}$\textbf{%
\ }and\textbf{\ }$\left| \mathrm{GHZ}\right\rangle ^{-}$\textbf{\ }can be
converted into each other through the local operation on any one of
intra-cavity qubits (say qubit\textbf{\ }$j_1$\textbf{): }$\left|
-\right\rangle _{j_1}\rightarrow -\left| -\right\rangle _{j_1}$\textbf{\ }and%
\textbf{\ }$\left| +\right\rangle _{j_1}\rightarrow \left| +\right\rangle
_{j_1}$. In this sense, the intra-cavity qubits can be prepared in a GHZ
entangled state deterministically.

As discussed here, a measurement on the coupler qubit is necessary to
prepare the intra-cavity qubits in a GHZ state. Note that fast and highly
accurate measurements on the state of a SC qubit are experimentally
available at this time (e.g., see [14]). In the following, we will propose
an alternative approach for entangling the intra-cavity qubits, which does
not require any measurement.

\begin{center}
\textbf{B. Method 2}
\end{center}

Assume that one intra-cavity SC qubit, say qubit 1$_1$ in cavity 1, is
initially in the state $\left| g^{\prime }\right\rangle ,$ each of all
remaining intra-cavity SC qubits is initially in the state $\left|
+\right\rangle ,$ and the coupler SC qubit $A$ is in the state $\alpha
\left| g^{\prime }\right\rangle _A+\beta \left| e\right\rangle _A$. Then the%
\emph{\ }initial state of the whole system is thus given by
\begin{equation}
\left| \mathrm{initial}\right\rangle =\left| g^{\prime }\right\rangle
_{1_1}\prod_{i=2}^{m_1}\left| +\right\rangle _{1_i}\otimes
\prod_{j=2}^N\prod_{i=1}^{m_j}\left| +\right\rangle _{j_i}\otimes \left(
\alpha \left| g^{\prime }\right\rangle _A+\beta \left| e\right\rangle
_A\right) .
\end{equation}
The procedure for preparing intra-cavity qubits in a GHZ entangled state is
listed as follows:

Step 1: Keep qubit $1_1$\ decoupled from cavity $1,$\ while adjust the level
spacings of other qubits such that their dynamics is governed by the
Hamiltonian~(6) (not including qubit $1_1$) for an interaction time $t$\
satisfying Eqs.~(9) and (10). By a similar derivation as shown in Eq.~(8),
one can easily find that the state (12) changes to
\begin{eqnarray}
&&\left| g^{\prime }\right\rangle _{1_1}\left( \alpha
\prod_{i=2}^{m_1}\left| +\right\rangle _{1_i}\otimes
\prod_{j=2}^N\prod_{i=1}^{m_j}\left| +\right\rangle _{j_i}\otimes \left|
g^{\prime }\right\rangle _A+\beta \prod_{i=2}^{m_1}\left| -\right\rangle
_{1_i}\otimes \prod_{j=2}^N\prod_{i=1}^{m_j}\left| -\right\rangle
_{j_i}\otimes \left| e\right\rangle _A\right) .  \nonumber \\
&&
\end{eqnarray}
Then, adjust the level spacings of the qubits such that the qubit system is
decoupled from the cavities.

Step 2: Perform the operations, $\left| g^{\prime }\right\rangle
_{1_1}\rightarrow \left| g\right\rangle _{1_1}$ and $\left| g^{\prime
}\right\rangle _A\rightarrow \left| g\right\rangle _A,$ by applying
classical pulses to qubit $1_1$ and the coupler qubit. In addition, perform
a swap operation $\left| g\right\rangle _{1_1}\left| e\right\rangle
_A\rightarrow -i\left| e\right\rangle _{1_1}\left| g\right\rangle _A$ [75],
which can be achieved by adjusting the level spacings of qubit $1_1$ and the
coupler qubit, such that the transitions $\left| g\right\rangle \rightarrow
\left| e\right\rangle $ of qubits $1_1$ and the coupler qubit are
dispersively coupled to cavity $1$ with the same detuning. Then the state
(13) becomes
\begin{eqnarray}
\left|\textrm{GHZ}\right\rangle \left| g\right\rangle _A &=&\left(
\alpha \left| g\right\rangle _{1_1}\prod_{i=2}^{m_1}\left| +\right\rangle
_{1_i}\otimes \prod_{j=2}^N\prod_{i=1}^{m_j}\left| +\right\rangle
_{j_i}-i\beta \left| e\right\rangle _{1_1}\prod_{i=2}^{m_1}\left|
-\right\rangle _{1_i}\otimes \prod_{j=2}^N\prod_{i=1}^{m_j}\left|
-\right\rangle _{j_i}\right) \left| g\right\rangle _A,  \nonumber \\
&&
\end{eqnarray}
which shows that all intra-cavity qubits are deterministically prepared in a
GHZ entangled state and disentangled from the coupler qubit. Since no
measurement is involved, the operations above are unitary. The operation
sequence for preparing the GHZ state is illustrated in Fig.~5.

\begin{figure}[tbp]
\begin{center}
\includegraphics[bb=27 23 582 442, width=10.0 cm, clip]{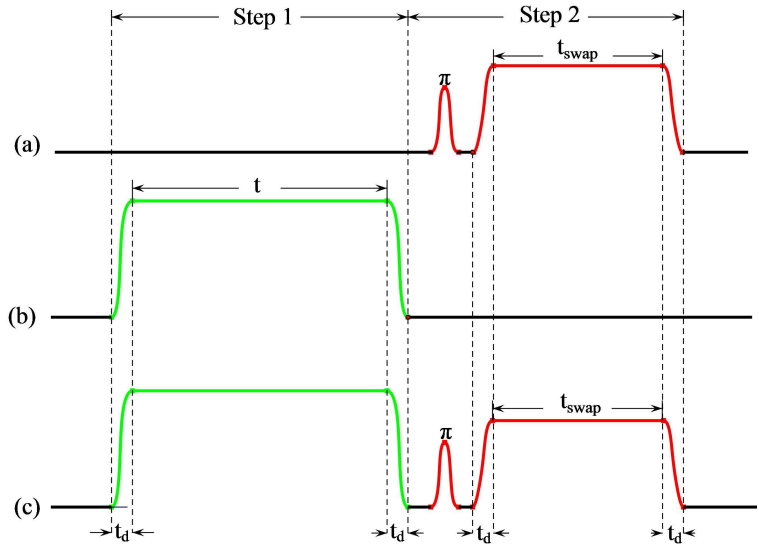} \vspace*{%
-0.08in}
\end{center}
\caption{(Color online) (a) Sequence of operations on qubit $1_1$. (b)
Sequence of operations on the intra-cavity qubits (except for qubit $1_1$). (c)
Sequence of operations on the coupler qubit $A$. The green curves in (b) and
(c) correspond to the operation described by step 1, which is for preparing
the state (13). The red curves in (a) and (c) correspond to the operation of
step 2 for preparing the state (14). The $\protect\pi$ pulse in (a)
transforms the state $\left| g^{\prime }\right\rangle _{1_1}\rightarrow
\left| g\right\rangle _{1_1}$, while the $\protect\pi$ pulse in (c)
transforms the state $\left| g^{\prime }\right\rangle _A\rightarrow \left|
g\right\rangle _A$. The right-hand square curves in (a) and (c) represent a
swap operation described by $\left| g\right\rangle _{1_1}\left|
e\right\rangle _A\rightarrow -i\left| e\right\rangle _{1_1}\left|
g\right\rangle _A$ [75]. In (a) and (c), $t_{\mathrm{swap}}$ is the swap operation
time. Note that the operations in (a), (b), and (c) are performed from left
to right.}
\end{figure}

From the description given above, one can see that the entanglement
production does not employ cavity photons. In addition, the GHZ state
preparation here does not depend on the number of intra-cavity qubits, which
requires only a few basic operations. Hence, the methods presented here for
entangling the intra-cavity qubits are quite simple.

\section{Experimental feasibility of entangling multiple qubits: an example}

To illustrate the experimental feasibility of our scheme, we consider a
system of two cavites (i.e., two one-dimensional transmission line
resonators) each hosting $M$ superconducting flux qubits (with $M\leqslant 4$%
)\ and coupled by a superconducting flux qubit $A.$ Figure~6 shows the setup
for each cavity hosting four flux qubits. The three levels $\left|
g\right\rangle ,$ $\left| g^{\prime }\right\rangle ,$ and $\left|
e\right\rangle $ of each flux qubit are depicted in Fig.~7. The $\left|
g\right\rangle \leftrightarrow \left| g^{\prime }\right\rangle $ transition
of each flux qubit can be made weak by increasing the potential barrier and
thus its coupling with the cavities is negligible. In reality, the $\left|
g^{\prime }\right\rangle \leftrightarrow \left| e\right\rangle $\textbf{\ }%
transition needs to be considered because the coupling between this
transition and each cavity may turn out to affect the operation fidelity.
Note that the $\left| g^{\prime }\right\rangle \leftrightarrow \left|
e\right\rangle $\textbf{\ }transition is much weaker than the $\left|
g\right\rangle \leftrightarrow \left| e\right\rangle $\textbf{\ }transition
due to the potential barrier between the two wells.

\begin{figure}[tbp]
\begin{center}
\includegraphics[bb=78 339 498 688, width=10.0 cm, clip]{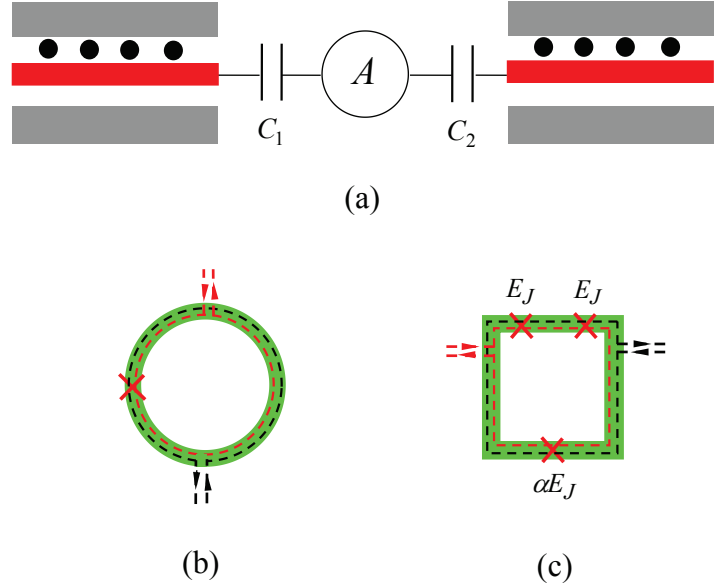} \vspace*{%
-0.08in}
\end{center}
\caption{(Color online) (a) Setup of two one-dimensional transmission line
resonators each hosting four flux qubits (dark dots) and coupled to flux
qubit $A$ (in the middle circle). The coupler qubit $A$ is connected to the
two resonators via capacitors $C_1$ and $C_2$, respectively. Each qubit
could be a radio-frequency superconducting quantum interference device (rf
SQUID) consisting of one Josephson junction enclosed by a superconducting
loop as depicted in (b), or a superconducting device with three Josephson
junctions enclosed by a superconducting loop as shown in (c). $E_J$ is the
Josephson junction energy ($0.6<\protect\alpha <0.8$). The superconducting
loop of each qubit, which is a large circle for (b) while a large square for
(c), is located in the plane of the resonators between the two lateral
ground planes. Each intra-cavity qubit is coupled to its cavity via the
magnetic flux through the superconducting loop of each qubit, which is
created by the cavity magnetic field threading the superconducting loop. The
intra-cavity qubits are placed at locations where the cavity magnetic fields
are the same to achieve an identical coupling strength for each qubit.
For each qubit, a tunable-coupler dc current line, e.g., the red dashed line in
(b) or (c) placed on the qubit loop, creates a dc magnetic pulse threading
the loop of each qubit, which is used for tuning the qubit level spacings.
Note that the qubit level spacings can be tuned by varying the magnitude of
the dc magnetic pulse through changing the current intensity. In addition,
for each qubit, a microwave bias ac current line, e.g., the dark dashed line in
(b) or (c) on the qubit loop, creates an ac magnetic pulse threading the
loop of each qubit, which is used to prepare the initial state of each qubit
or/and manipulate the state of each qubit during the GHZ state preparation.}
\label{fig:6}
\end{figure}

\begin{figure}[tbp]
\begin{center}
\includegraphics[bb=43 219 573 444, width=12.5 cm, clip]{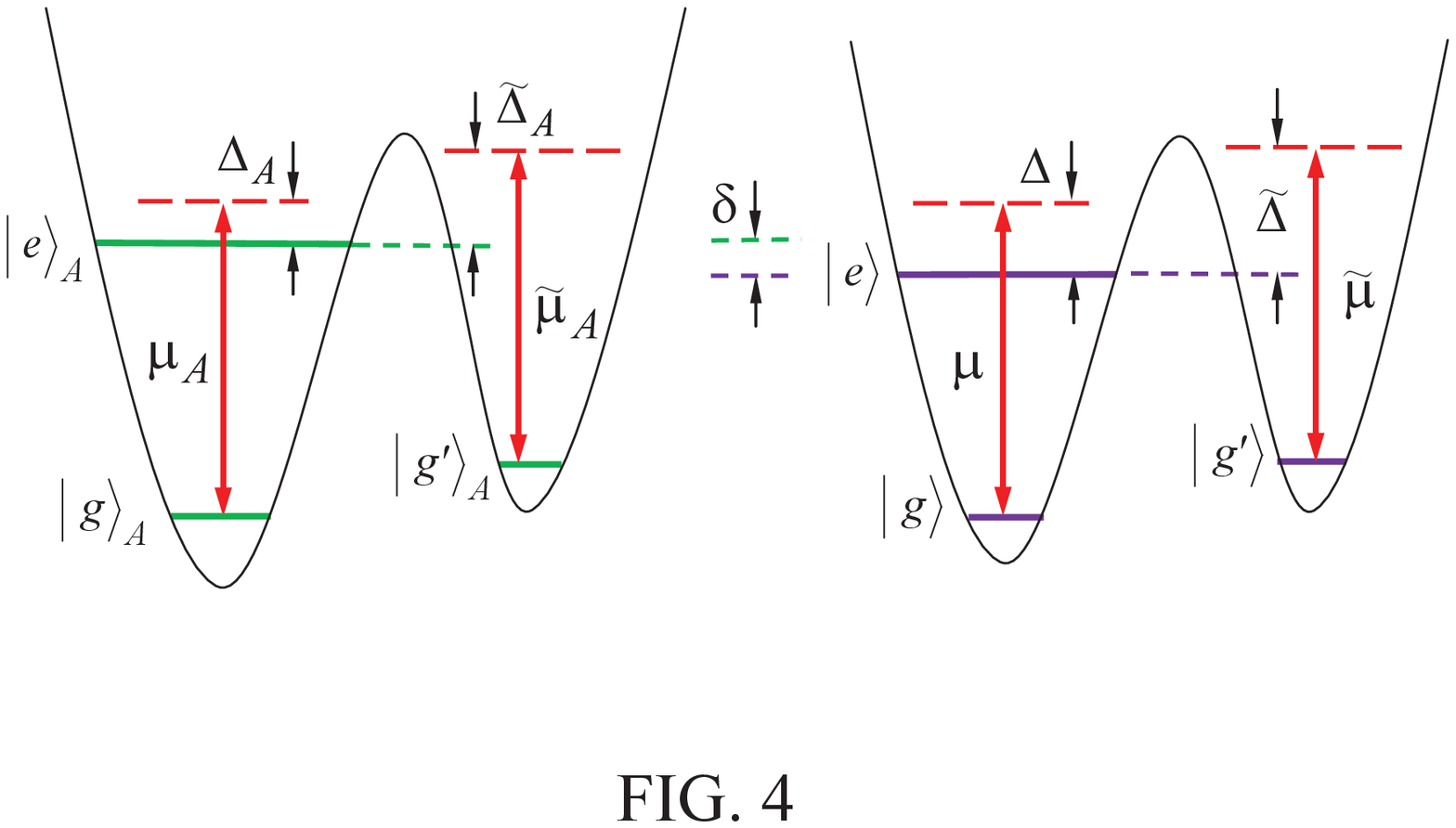} \vspace*{%
-0.08in}
\end{center}
\caption{(Color online) Illustration of the qubit-cavity dispersive
interaction. The left is for the coupler flux qubit $A$ , while the right is
for the intra-cavity flux qubits. The tunneling between the two lowest
levels is made weak by increasing the potential barrier, such that the level
$\left\vert g^{\prime }\right\rangle$ can be stored for a long time.}
\label{fig:7}
\end{figure}

When the unwanted coupling of the $\left| g^{\prime }\right\rangle
\leftrightarrow \left| e\right\rangle $ transition with the cavities and the
unwanted inter-cavity crosstalk are included, the Hamiltonian (1) is
modified as
\begin{eqnarray}
\widetilde{H} &=&\sum_{j=1}^2\mu \left( e^{-i\Delta t}a_jJ_j^{+}+\mathrm{H.c.%
}\right) +\sum_{j=1}^2\mu _A\left( e^{-i\Delta _At}a_j\sigma _A^{+}+\mathrm{%
H.c.}\right)  \nonumber \\
&&+\sum_{j=1}^2\widetilde{\mu }\left( e^{-i\widetilde{\Delta }t}a_j%
\widetilde{J}_j^{+}+\mathrm{H.c.}\right) +\sum_{j=1}^2\widetilde{\mu }%
_A\left( e^{-i\widetilde{\Delta }_At}a_j\widetilde{\sigma }_A^{+}+\mathrm{%
H.c.}\right)  \nonumber \\
&&+\mu _{12}\left( a_1a_2^{+}+\mathrm{H.c.}\right) ,
\end{eqnarray}
where $J_j^{+}=\sum_{j=1}^2\sum_{i=1}^M\sigma _{j_i}^{+},$ $\widetilde{J}%
_j^{+}=\sum_{j=1}^2\sum_{i=1}^M\widetilde{\sigma }_{j_i}^{+},$ $\widetilde{%
\sigma }_{j_i}^{+}=\left| e\right\rangle _{j_i}\left\langle g^{\prime
}\right| ,$ and $\widetilde{\sigma }_A^{+}=\left| e\right\rangle
_A\left\langle g^{\prime }\right| .$ The terms in the first (second) pair of
parentheses in the second line describe the unwanted coupling between the%
\textbf{\ }$\left| g^{\prime }\right\rangle \leftrightarrow \left|
e\right\rangle $\textbf{\ }transition of intra-cavity qubits (coupler qubit)
and the respective cavities with coupling strength\textbf{\ }$\widetilde{\mu
}$\textbf{\ (}$\widetilde{\mu }_A$\textbf{) }and detuning\textbf{\ }$%
\widetilde{\Delta }=\omega _c-\omega _{eg^{\prime }}$\textbf{\ (}$\widetilde{%
\Delta }_A=\omega _c-\omega _{eg^{\prime },A}$\textbf{). }The terms in the
last pair of parentheses\textbf{\ }describe the inter-cavity crosstalk
between two cavities, with the inter-cavity coupling constant $\mu _{12}$.

The dynamics of the lossy system, with finite qubit relaxation, dephasing
and photon lifetime being included, is determined by the following master
equation
\begin{equation}
\frac{d\rho }{dt}=-i\left[ \widetilde{H},\rho \right] +\sum_{j=1}^2\kappa _j%
\mathcal{L}\left[ a_j\right] +P+Q,
\end{equation}
with
\begin{eqnarray}
P &=&\sum_{j=1}^2\sum_{i=1}^M\left\{ \gamma _{_{eg}}\mathcal{L}\left[ \sigma
_{j_i}\right] +\gamma _{eg^{\prime }}\mathcal{L}\left[ \widetilde{\sigma }%
_{j_i}\right] +\gamma _{g^{\prime }g}\mathcal{L}\left[ \overline{\sigma }%
_{j_i}\right] \right\}  \nonumber \\
&&\ +\sum_{j=1}^2\sum_{i=1}^M\left\{ \gamma _{e,\varphi }\left( E_{j_i}\rho
E_{j_i}-E_{j_i}\rho /2-\rho E_{j_i}/2\right) \right\}  \nonumber \\
&&\ +\sum_{j=1}^2\sum_{i=1}^M\left\{ \gamma _{g^{\prime },\varphi }\left(
G_{j_i}^{\prime }\rho G_{j_i}^{\prime }-G_{j_i}^{\prime }\rho /2-\rho
G_{j_i}^{\prime }/2\right) \right\} ,
\end{eqnarray}
\begin{eqnarray}
Q &=&\gamma _{_{eg},A}\mathcal{L}\left[ \sigma _A^{-}\right] +\gamma
_{eg^{\prime },A}\mathcal{L}\left[ \widetilde{\sigma }_A^{-}\right] +\gamma
_{g^{\prime }g,A}\mathcal{L}\left[ \overline{\sigma }_A^{-}\right]  \nonumber
\\
&&\ +\gamma _{e,\varphi ,A}\left( E_A\rho E_A-E_A\rho /2-\rho E_A/2\right)
\nonumber \\
&&\ +\gamma _{g^{\prime },\varphi ,A}\left( G_A^{\prime }\rho G_A^{\prime
}-G_A^{\prime }\rho /2-\rho G_A^{\prime }/2\right) ,
\end{eqnarray}
where $\overline{\sigma }_{j_i}=\left| g\right\rangle _{j_i}\left\langle
g^{\prime }\right| ,\overline{\sigma }_A=\left| g\right\rangle
_A\left\langle g^{\prime }\right| ;$ $E_{j_i}=\left| e\right\rangle
_{j_i}\left\langle e\right| ,G_{j_i}^{\prime }=\left| g^{\prime
}\right\rangle _{j_i}\left\langle g^{\prime }\right| ,G_A^{\prime }=\left|
g^{\prime }\right\rangle _A\left\langle g^{\prime }\right| ;$ and $\mathcal{L%
}\left[ \Lambda \right] =\Lambda \rho \Lambda ^{+}-\Lambda ^{+}\Lambda \rho
/2-\rho \Lambda ^{+}\Lambda /2,$ with $\Lambda =a_j,\sigma _{j_i},\widetilde{%
\sigma }_{j_i},\overline{\sigma }_{j_i},\sigma _A,\widetilde{\sigma }_A,%
\overline{\sigma }_A.$ Here, $\kappa _j$ is the photon decay rate of cavity $%
a_j$. In addition, $\gamma _{g^{\prime }g}$ is the energy relaxation rate of
the level $\left| g^{\prime }\right\rangle $ of qubits, $\gamma _{eg}$ ($%
\gamma _{eg^{\prime }}$) is the energy relaxation rate of the level $\left|
e\right\rangle $ of intra-cavity qubits for the decay path $\left|
e\right\rangle \rightarrow \left| g\right\rangle $ ($\left| g^{\prime
}\right\rangle $), and $\gamma _{e,\varphi }$ ($\gamma _{g^{\prime },\varphi
}$) is the dephasing rate of the level $\left| e\right\rangle $ ($\left|
g^{\prime }\right\rangle $) of intra-cavity qubits. The symbols\ $\gamma
_{g^{\prime }g,A},$\ $\gamma _{eg,A},$\ $\gamma _{eg^{\prime },A},$\ $\gamma
_{e,\varphi ,A},$\ and $\gamma _{g^{\prime },\varphi ,A}$\ denote the
corresponding decoherence rates of the coupler qubit $A.$

The fidelity of the operation is given by [79]

\begin{equation}
\mathcal{F}=\sqrt{\left\langle \psi _{\mathrm{id}}\right| \rho \left| \psi _{%
\mathrm{id}}\right\rangle },
\end{equation}
where $\left| \psi _{\mathrm{id}}\right\rangle $ is the output state of an
ideal system (i.e., without dissipation, dephasing, and crosstalks) as
discussed in the previous section, and $\rho $ is the final density operator
of the system when the operation is performed in a realistic physical
system. Without loss of generality, consider now that the qubit system is
initially in the state $\prod_{j=1}^2\prod_{i=1}^M\left| +\right\rangle
_{j_i}\left( \left| g^{\prime }\right\rangle _A+\left| e\right\rangle
_A\right) /\sqrt{2}$ (with $M\leq 4$), and the two cavities are initially in
a vacuum state, for which the ideal state $\left| \psi _{\mathrm{id}%
}\right\rangle $ is the one given in Eq.~(11) with $\alpha =\beta =1/\sqrt{2}
$.

\begin{figure}[tbp]
\begin{center}
\includegraphics[bb=10 62 860 599, width=15.0 cm, clip]{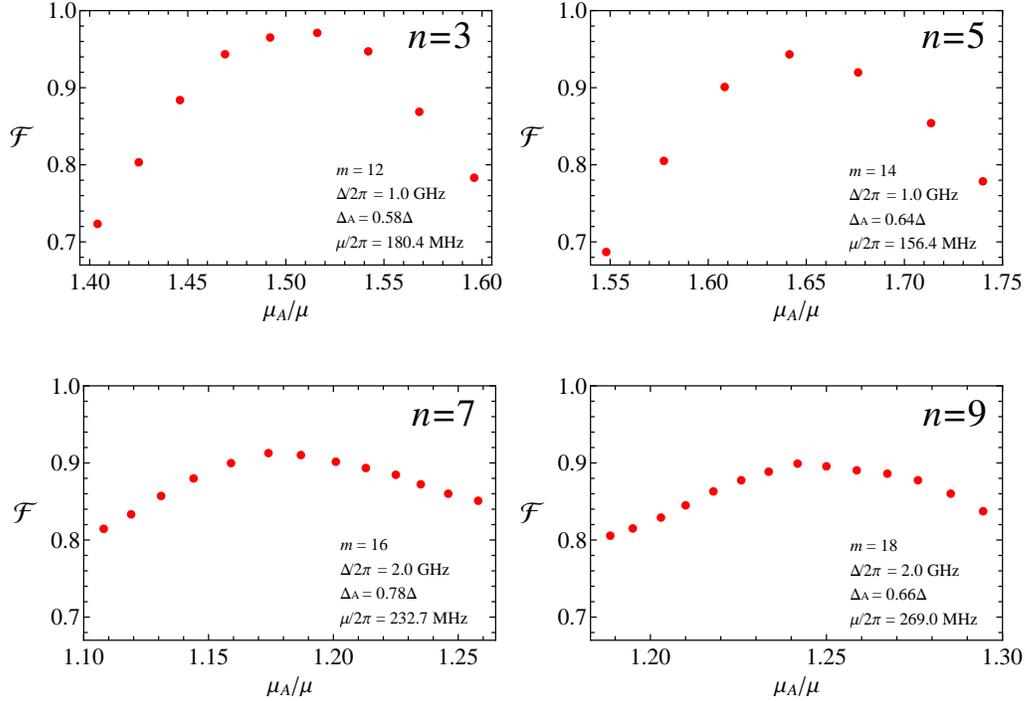} \vspace*{%
-0.08in}
\end{center}
\caption{(Color online) Fidelity versus $\protect\mu_A /\protect\mu$. Here, $%
n$ is the total number of qubits including the coupler qubit. Also $n=3, 5,
7 $ and $9$ correspond to the cases of each cavity hosting one, two, three
and four qubits, respectively. To have an achievable good fidelity with
reasonable parameters, we increase $\Delta/(2\protect\pi)=2$ from 1 to 2
GHz, when going from 5 to 7 qubits. For $n=3,$ $5,$ $7$ and $9,$ a high
fidelity $\sim 97.2\%,$ $94.3\%,$ $91.3\%,$ and $90.1\%$ can be achieved
with $\protect\mu _A/\protect\mu $ being $1.516,$ $1.641,$ $1.174,$ and $%
1.242$, respectively.}
\label{fig:8}
\end{figure}

We consider identical intra-cavity flux qubits. Given $\Delta ,$ $\Delta _A,$
and $m,$ the coupling constant $\mu $ is determined by $\mu =\sqrt{2N\left|
\delta \right| \Delta _A/m}\Delta /\left( \Delta +\Delta _A\right) $
[derived from Eqs.~(9) and (10)]. We set $\widetilde{\mu }=0.1\mu $ and $%
\widetilde{\mu }_A=0.1\mu _A,$ which is a good approximation by increasing
the potential barrier such that the transition matrix element between the
two levels $\left| g^{\prime }\right\rangle $ and $\left| e\right\rangle $
is smaller than that between the two levels $\left| g\right\rangle $ and $%
\left| e\right\rangle $ by one order of magnitude (Fig.~7). We choose $%
\Delta /\left( 2\pi \right) =1.0$ GHz for $n=3$ and $n=5$, while $\Delta
/\left( 2\pi \right) =2.0$ GHz for $n=7 $ and $n=9$. Here,\emph{\ }$n=2M+1$%
\emph{\ }is the total number of qubits including the coupler qubit. In
addition, we set $\widetilde{\Delta }=\Delta +2\pi \times 1.5$ GHz and $%
\widetilde{\Delta }_A=\Delta _A+2\pi \times 1.5$ GHz. We now choose $\kappa
_j^{-1}=15$ $\mu $s, $\gamma _{e,\varphi }^{-1}=5$ $\mu $s, $\gamma
_{g^{\prime },\varphi }^{-1}=7.5$ $\mu $s, $\gamma _{eg}^{-1}=5$ $\mu $s, $%
\gamma _{eg^{\prime }}^{-1}=7.5$ $\mu $s, and $\gamma _{g^{\prime
}g}^{-1}=15 $ $\mu $s (a conservative consideration, e.g., see Ref.~[9]).
Our numerical calculations show that when $\mu _{12}$ is smaller than $\mu
_A $ by two orders of magnitude, the effect of the inter-cavity crosstalk on
the operation is negligible. Thus, we now set $\mu _{12}=0.01\mu _A.$ With
the parameters chosen here and by numerically optimizing the parameters $m,$
$\mu ,$ and $\Delta _A$, the fidelity versus $\mu _A/\mu $ is plotted in
Fig.~8 for $n=3,$ $5,$ $7$ and $9.$ From Fig.~8, one can see that for $n=3,$
$5,$ $7$ and $9,$ a high fidelity $\sim 97.2\%,$ $94.3\%,$ $91.3\%,$ and $%
90.1\%$ can be achieved with the optimized values of $\mu _A/\mu $ being $%
1.516,$ $1.641,$ $1.174,$ and $1.242$, respectively. We remark that the
fidelity can be further increased by improving the system parameters.

Figure~8 shows that with the parameter values chosen above the large
detuning condition is not well satisfied for the optimized fidelity, e.g., $%
\Delta _A/\mu _A\sim 3.95$\ for $n=9$, which implies that the state
evolution determined by the effective Hamiltonian~(6) can be a good
approximation with a suitable choice of parameters, even when the
qubit-cavity detunings are not much larger than the coupling strengths. This
result has a quantitative explanation. Beyond the large detuning regime,
when the coupler qubit is initially in the state $\left| e\right\rangle $,
the total cavity-qubit system undergoes Rabi oscillations in the
corresponding single-excitation subspace. The associated Rabi frequencies
have a dependence on the number of the intracavity qubits in the state $%
\left| g\right\rangle $. With a suitable choice of the qubit-cavity coupling
strengths and detunings, all of the state components with the coupler qubit
being initially in $\left| e\right\rangle $\ can return to their initial
forms almost at the same time, with the resulting phase shift being related
to the corresponding Rabi frequency.

As discussed in [39,80], the condition $\mu _{12}\leq 0.01\mu _A$ can be met
with the typical capacitive cavity-qubit coupling. Figure~8 shows that at
the optimum points, the coupling strengths are\ $\left\{ \mu /2\pi ,\mu
_A/2\pi \right\} \sim \{180.4$ MHz, $273.5$ MHz$\}$ ($n=3$)$,$ $\{156.4$
MHz, $256.75$ MHz$\}$ ($n=5$)$,$ $\{232.7$ MHz, $273.2$ MHz$\}$ ($n=7$)$,$
and $\{269.0$ MHz, $334.0$ MHz$\}$ ($n=9$)$.$ The coupling strengths of
these values are readily achievable in experiments because a coupling
strength $\sim 636$ MHz has been reported for a superconducting flux device
coupled to a one-dimensional transmission line resonator [26]. For a flux
qubit, the $\left| g\right\rangle \leftrightarrow \left| e\right\rangle $
transition frequency could be between 5 GHz and 20 GHz. Thus, we\textbf{\ }%
can\textbf{\ }choose\textbf{\ }$\omega _c/2\pi \sim 7.5$ GHz. For the value
of $\kappa _j^{-1}$ used in the numerical calculation, the required quality
factor for each cavity is $Q=\kappa _j^{-1}\omega _c\sim 7.1\times 10^5,$
which is available in experiments according to previous reports [81,82].
Therefore, the high-fidelity creation of GHZ states of up to nine qubits by
using this proposal is feasible\textbf{\ }with current circuit QED
technology.

\begin{table}[tbp]
\begin{center}
\includegraphics[bb=81 532 542 654, width=15.0 cm, clip]{table1.eps}
\vspace*{-0.18in}
\end{center}
\caption{Qubits $1_1$, $1_2$, $1_3$, and $1_4$ are four qubits placed in one
cavity; while Qubits $2_1$, $2_2$, $2_3$, and $2_4$ are four qubits placed
in the other cavity. Here $\protect\mu$ takes the same value used in Fig. 8.}
\label{table:1}
\end{table}

\begin{table}[tbp]
\begin{center}
\includegraphics[bb=83 571 551 700, width=15.0 cm, clip]{table2.eps}
\vspace*{-0.18in}
\end{center}
\caption{Here $\Delta$ takes the same value used in Fig. 8.}
\label{table:2}
\end{table}

To see how well this method works in a more realistic situation, we now
consider inhomogeneous coupling of qubits to the mode in each cavity,
non-uniform distribution of qubit frequencies, imperfect preparation of the
initial states, error in the operation time, and the existence of thermal
photons in each cavity. The density operator of a thermal state of each
cavity is described by $\rho =\sum\limits_{n=0}^{\infty }\frac{\overline{n}%
^{n}}{\left( 1+\overline{n}\right) ^{n+1}}\left\vert n\right\rangle
\left\langle n\right\vert ,$\ with $\overline{n}$\ being the average photon
number and $\left\vert n\right\rangle $\ being an $n$-photon state.\ In our
numerical simulation, we choose $\overline{n}=0.1$. The initial state of the
qubit system is modified as $\prod_{j=1}^{2}\prod_{i=1}^{M}\left( \sqrt{%
\frac{1+\varepsilon }{2}}\left\vert g^{\prime }\right\rangle _{j_{i}}+\sqrt{%
\frac{1-\varepsilon }{2}}\left\vert g\right\rangle _{j_{i}}\right) \left(
\sqrt{\frac{1+\varepsilon }{2}}\left\vert g^{\prime }\right\rangle _{A}+%
\sqrt{\frac{1-\varepsilon }{2}}\left\vert e\right\rangle _{A}\right) .$\ For
simplicity, we here consider the identical error $\varepsilon $\ for the
preparation of the initial state of each qubit. Without loss of generality,
we will numerically investigate how the maximum fidelity in each subfigure
of Fig.~8 is affected by the above-mentioned factors.

As mentioned above, the coupling constant $\mu _{A}$\ of the coupler qubit,
corresponding to the maximum fidelity in each subfigure of Fig. 8, is
calculated to be $\mu _{A}=1.516\mu $\ for $n=3$; $\mu _{A}=1.641\mu $\ for $%
n=5$; $\mu _{A}=1.174\mu $\ for $n=7$; and $\mu _{A}=1.242\mu $\ for $n=9$.
Here, the value of $\mu $\ is shown in Fig. 8. In addition, the detuning $%
\Delta _{A}$\ for the coupler qubit takes the same value as in Fig. 8.

The coupling constants and the detunings for the intra-cavity qubits are
listed in Tables 1 and 2, where up to $1\%$\ inhomogeneous coupling
constants and up to $5\%$\ non-uniform detunings are considered. In Tables 1
and 2, the values of $\mu $\ and $\Delta $\ are the same as those in Fig. 8.

With the parameters chosen above, in Fig. 9 we present a numerical
simulation of the fidelity versus $t/T$\ for $\varepsilon =0,$\ $0.05,$\ and
$0.1.$\ Here, $t$\ is the operation time, while $T$\ is the optimal
operation time corresponding to the maximum fidelities in Fig. 8, which are
46.55 ns, 67.94 ns, 167.19 ns, 106.43 ns for $n=3,5,7,$\ and $9$,
respectively. From Fig. 9, one can see that the fidelity is insensitive to
the error $\varepsilon $\ but is significantly affected by the error in
operation time. For $t=0.95T$\ or $1.05T$\ (i.e., $5\%$\ operational time
error), the fidelity drops down to $\lesssim 0.5.$\ Note that for $t=T,$\
good fidelities $\gtrsim 0.939$\ for $n=3$, $\gtrsim 0.845$\ for $n=5$, $%
\gtrsim 0.778$\ for $n=7$, and $\gtrsim 0.760$\ for $n=9$\ can be obtained.

\begin{figure}[tbp]
\begin{center}
\includegraphics[bb=1 54 1007 766, width=15.0 cm, clip]{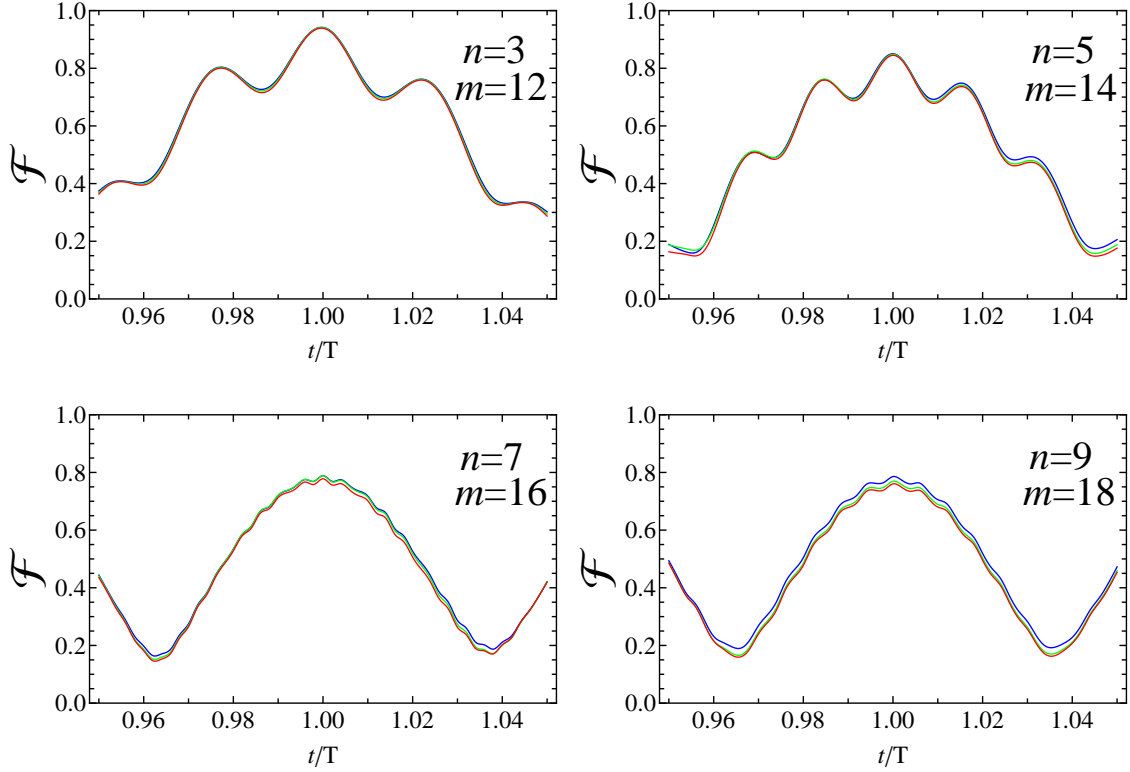} \vspace*{%
-0.08in}
\end{center}
\caption{(Color online) Fidelity versus $t/T$. Here, $t$ is the operation
time, while $T$ is the operation time corresponding to the maximum fidelities
in Fig.~5, which are 46.55 ns, 67.94 ns, 167.19 ns, 106.43 ns for $n=3, 5,
7, $ and $9$, respectively. The blue, green, and red curves correspond to $%
\protect\varepsilon =0, 0.05$, and $0.1$. For $t=T$, the fidelities
corresponding to $\protect\varepsilon =\left\{ 0, 0.05, 0.1\right\} $ are: $%
\left\{0.942, 0.941, 0.939\right\} $ for $n=3$; $\left\{0.850, 0.848,
0.845\right\} $ for $n=5$; $\left\{0.790, 0.789, 0.778\right\} $ for $n=7$;
and $\left\{0.786, 0.769, 0.760\right\} $ for $n=9$.}
\label{fig:9}
\end{figure}

It is worthwhile to discuss the advantage of utilizing negative detunings
versus positive detunings. For the flux qubits with three levels $\left\vert
g\right\rangle $, $\left\vert g^{\prime }\right\rangle $, and $\left\vert
e\right\rangle $\ shown in Fig. 7, the purpose of using a negative detuning
is to increase the detuning of the $\left\vert g^{\prime }\right\rangle
\leftrightarrow \left\vert e\right\rangle $\ transition frequency from the
cavity frequency, in order to reduce the effect of this unwanted transition
on the operation fidelity. It can be seen from Fig. 7 that the detuning of
the $\left\vert g^{\prime }\right\rangle \leftrightarrow \left\vert
e\right\rangle $\ transition frequency from the cavity frequency would be
smaller when using a positive detuning (i.e, the case when the cavity
frequency is smaller than the $\left\vert g^{\prime }\right\rangle
\leftrightarrow \left\vert e\right\rangle $\ transition frequency), compared
to using the negative detuning.\emph{\ }

\section{Conclusion}

We have proposed a general and efficient way to entangle SC qubits in a
multi-cavity system. In principle, GHZ states of \textit{an arbitrary number}
of intra-caviy qubits plus the coupler qubit can be created through a single
operation and without any measurements. Since only virtual photon processes
take place, the decoherence caused by cavity decay and the effect of
unwanted inter-cavity crosstalk are greatly suppressed. Also, the higher
energy level is not occupied for any intra-cavity qubit; thus decoherence
from the qubits is much reduced. In addition, we have introduced two simple
methods for entangling the intra-cavity qubits in a GHZ state. Our numerical
simulations show that it is feasible to generate high-fidelity GHZ entangled
states with up to nine SC qubits in a circuit consisting of two resonators.
We hope this will stimulate future experimental activities. The method
presented here is quite general and can be applied to various other physical
systems. We believe that the cluster-style architecture shown in Fig. 1 has
applications in fault-tolerant code for scalable quantum computing. Multiple
physical qubits in each cavity can be used to construct a logic qubit, as
required by error-correction protocols. The system can also be used to
simulate the dynamics of the star-type coupled spin system, where many spins
are coupled to a common spin.

\begin{center}
{\bf ACKNOWLEDGMENTS}
\end{center}

We very gratefully acknowledge Dr.~Anton Frisk Kockum for a critical
reading of the manuscript. This work was partly supported by the RIKEN iTHES
Project, the MURI Center for Dynamic Magneto-Optics via the AFOSR award
number FA9550-14-1-0040, and a Grant-in-Aid for Scientific Research (A). It
was also partially supported by the Major State Basic Research Development
Program of China under Grant No.~2012CB921601, the National Natural Science
Foundation of China under Grant Nos.~[11074062,~11374083], the Zhejiang
Natural Science Foundation under Grant No.~LZ13A040002, the funds of
Hangzhou Normal University under Grant Nos.~[HSQK0081,~PD13002004], and the
funds of Hangzhou City for supporting the Hangzhou-City Quantum Information
and Quantum Optics Innovation Research Team.

\end{document}